\documentclass[twocolumn,showpacs,preprintnumbers,amsmath,amssymb]{revtex4}
\usepackage{graphicx}
\usepackage{dcolumn}
\usepackage{bm}
\begin{document}
\title{Emergent Collectivity in Nuclei and Enhanced Proton-Neutron Interactions}
\author{D.~Bonatsos$^{1}$, S.~Karampagia$^{1}$, R.B.~Cakirli$^{2,3}$, R.F.~Casten$^{4}$, K.~Blaum$^{3}$, L.~Amon Susam$^{2}$}
\affiliation{$^1$Institute of Nuclear and Particle Physics,
National Centre for Scientific Research Demokritos, GR-153 10
Aghia Paraskevi, Attiki, Greece} \affiliation{$^2$Department of
Physics, University of Istanbul, Istanbul, Turkey}
\affiliation{$^3$Max-Planck-Institut f\"{u}r Kernphysik,
Saupfercheckweg 1, D-69117 Heidelberg, Germany}
\affiliation{$^4$Wright Nuclear Structure Laboratory, Yale
University, New Haven, CT 06520, USA}
\date{\today}
\pacs{21.60.Cs, 21.10.Dr, 21.60.Ev, 21.30.Fe}
\begin{abstract}
Enhanced proton-neutron interactions occur in heavy nuclei along a
trajectory of approximately equal numbers of valence protons and
neutrons. This is also closely aligned with the trajectory of the
saturation of quadrupole deformation. The origin of these enhanced
$p$-$n$ interactions is discussed in terms of spatial overlaps of
proton and neutron wave functions that are orbit-dependent. It is
suggested for the first time that nuclear collectivity is driven
by synchronized filling of protons and neutrons with orbitals
having parallel spins, identical orbital and total angular momenta
projections, belonging to adjacent major shells and differing by
one quantum of excitation along the $z$-axis. These results may
lead to a new approach to symmetry-based theoretical calculations
for heavy nuclei.
\end{abstract}

\maketitle

\section{Introduction}
In many areas of science, coherence and correlations emerge in
complex many-body systems from microscopic ingredients and their
interactions. Examples abound in the vibrational, rotational, and
bending modes of atoms and molecules and in spatial patterns in
complex molecules \cite{Iachello}, in collectivity and phase
transitions in atomic nuclei and in similarities to correlations
in cold atoms \cite{Zinner}. Similar physics appears in pattern
formation in biological entities (e.g., Turing model
\cite{Turing}), cooperativity in biochemical signaling
\cite{Bialek}, in self-organized social behavior in animal
species, in ecological environments \cite{Fossion1,Fossion2}, and
in climatic tipping points \cite{Lenton}. The over-arching
question cutting across disciplines is how assemblages of
interacting constituents can develop emerging collectivity not
apparent in the individual constituents.

Atomic nuclei provide a fascinating venue for such studies. Their
structure is primarily determined by two forces (strong and
Coulomb) whose relative strengths are proton number dependent.
Further, one can often control the number of interacting bodies
(nucleons) and study the particle-number dependence of collective
phenomena. Studying how the often simple behavior of nuclei can
emerge from nucleonic interactions has been described as one of
the great challenges in the study of nuclei \cite{NA,LRP}. The key
residual interactions are those among the valence nucleons, and,
in particular, the residual valence proton-neutron ($p$-$n$)
interactions \cite{deShalit,Talmi,F-P,Casten,Otsuka}.

It is the purpose of this paper to, first, show newly discovered
singular aspects of $p$-$n$
 interactions in nuclei with equal or
nearly equal numbers of \textit{valence} protons and neutrons and,
secondly, to relate these enhanced interactions to the onset of
collectivity. We will then exploit an empirical relation between
the single particle quantum numbers of the last-filled proton and
neutron orbitals in these nuclei to suggest a simple
interpretation of those $p$-$n$ interactions in terms of spatial
overlaps of their wave functions. Finally, we show that the nearly
synchronous filling of such pairs of orbitals correlates well with
the growth and saturation of collectivity. This leads to a
suggestion for a possible new coupling scheme that could greatly
simplify symmetry-based shell model calculations.

\begin{figure*}
\includegraphics[width=1\textwidth]{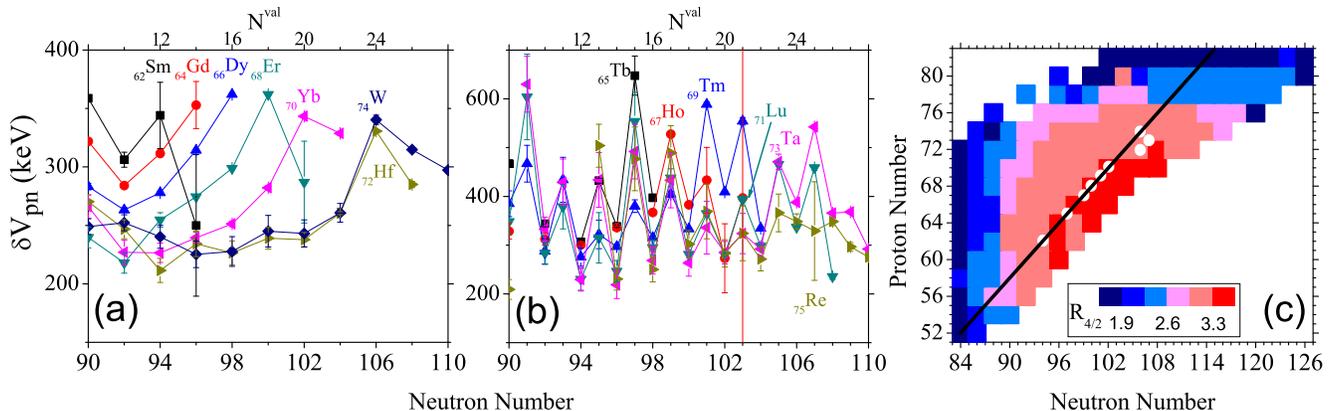}\vspace{-5.5cm}
\caption{(Color online) Empirical $\delta$$V_{pn}$ values for
even-$Z$ nuclei (Based on Ref.~\cite{Cakirli2}) (a). Empirical
$\delta$$V_{pn}$ values for odd-$Z$ nuclei (b). Color coded
contour plot of empirical $R_{4/2}$ values in the $Z$= 50-82, $N$=
82-126 shells (c). The line drawn represents the line of
$Z_{val}$=$N_{val}$. The white circles are the nuclides for each
$Z$ where the largest $\delta$$V_{pn}$ value is observed.}
\end{figure*}

\section{Empirical p-n Interactions}

A measure of the average $p$-$n$ interaction of the last nucleons
can be extracted from a double difference of binding energies,
called $\delta$$V_{pn}$ \cite{Jingye}. In
Refs.~\cite{Isacker,Cakirli1,Brenner,Oktem,Mario,Chen,Dennis1}
$\delta$$V_{pn}$ was related to shell effects and the onset of
deformation. In Ref.~\cite{Brenner1} it was shown that
$\delta$$V_{pn}$ has large singularities for light $Z$=$N$ nuclei
linked \cite{Isacker} to maximal spatial-spin overlaps of proton
and neutron wave functions.

One expects such a phenomenon to dissipate in heavier nuclei where
spin-orbit and Coulomb forces grow in importance. And, of course,
$Z$=$N$ nuclei do not exist beyond $A$$\sim$100. Thus it came as a
surprise that $\delta$$V_{pn}$ values in heavy nuclei show
similar, though highly muted, peaks \cite{Cakirli2}, as shown in
Fig.~1(a), when the number of $\textit{valence}$ neutrons equals
the number of $\textit{valence}$ protons or, late in the shells,
slightly exceeds the valence proton number. Interestingly, there
is a special quantal relation between the last-filled proton and
neutron Nilsson \cite{Nilsson} orbitals (these are all deformed
nuclei) in many nuclei exhibiting these singular $\delta$$V_{pn}$
values, namely, that these orbitals are often related by $\Delta
K$[$\Delta N$,$\Delta{n_z}$,$\Delta \Lambda$]=0[110],
 where $K$ and $\Lambda$
are the projections of the total and orbital angular momenta on
the $z$-axis ($K$=$\Lambda$$\pm$1/2), respectively. If both the
oscillator quantum number \textit{N} ($N$=${n_x}$+${n_y}$+${n_z}$)
and the number of quanta in the $z$-direction (the deformation
axis), $n_z$, increase by one, then $n_x$+$n_y$ is constant: the
two wave functions differ by a single quantum in the $z$-direction
and are therefore highly overlapping.

These results concern nuclei with even numbers of protons and the
peaks in $\delta$$V_{pn}$ were for even-even nuclei. It is well
known in such nuclei that the ground state wave functions are
spread out over several orbits due to the pairing force. Therefore
a much more direct and $\textit{pure}$ perspective is given by
odd-odd nuclei where the last protons and last neutrons occupy
specific single orbits. The panel (b) of Fig.~1 shows for the
first time the empirical results for $\delta$$V_{pn}$ for odd-$Z$
nuclei with both even and odd-$N$. Not only do these results also
show spikes, at $Z_{val}\simeq N_{val}$, but now the peaks are
sharper and greatly enhanced in magnitude (about 4 times larger
than for even-even nuclei). Figure 1(c) shows the locus of maximum
$\delta$$V_{pn}$ values in an $Z$-$N$ plot of $R_{4/2}$$\equiv$
E($4_1^+$)/E(2$_1^+$), which varies from $<$2 near closed shells
to $\sim$3.33 for well-deformed axial rotors. The results for
even-even and odd-odd nuclei closely match both the $Z_{val}\simeq
N_{val}$ line and the onset of deformation occuring for
$R_{4/2}$$>$ 3.3. This highlights the link to the evolution and
saturation of collectivity.

\section{A Simple Model for the p-n Interactions:Calculations and Comparison with Empirical Results}
How can one try to understand the origin and implications for
these results? One approach is large scale computationally
intensive methods such as Density Functional Theory calculations
which, indeed, were compared to empirical trends of
$\delta$$V_{pn}$ in Ref.~\cite{Mario}. While this approach yields
good agreement with the data it does not reveal per se the
underlying origin of the behavior of $\delta$$V_{pn}$. Here we
take a much simpler theoretical perspective by directly
calculating spatial overlaps of proton and neutron Nilsson wave
functions. Our approach in fact obtains similar results but now in
a way that explicitly exposes, in a physically intuitive way, the
underlying origins of the emergent collectivity through the roles
of specific orbitals in $p$-$n$ interactions. As will be seen,
this uncovers a heretofore unrecognized pattern in the synchronous
filling of proton and neutron orbitals that helps explain the
evolution of collectivity and its locus in $Z$ and $N$.

Nilsson wave functions in the form \cite{Nilsson} $\chi_{N\Omega}=
\sum_{l \Lambda} a_{l \Lambda}^\Omega | N l \Lambda \Sigma
\rangle$ were used, where $\Omega$, $\Lambda$, $\Sigma$ are the
projections of the total particle angular momentum $j$, the
orbital angular momentum $l$ and the spin $s$ on the $z$-axis,
while the coefficients $a_{l \Lambda}^\Omega$ were calculated by
solving the Nilsson Hamiltonian with the standard parameter
values, $\kappa$= 0.0637 and $\mu$= 0.42 for neutrons and 0.0637
and 0.6 for protons, respectively. For axially symmetric nuclei,
which we deal with here, $K$, the projection of the total angular
momentum on the $z$-axis, and $\Omega$ are the same. Overlaps
$\int ({\chi^*_{N_1 \Omega_1} \chi_{N_1 \Omega_1}}) ({\chi^*_{N_2
\Omega_2} \chi_{N_2 \Omega_2}}) dV$ were calculated using
spherical coordinates. Though the deformation dependence is weak,
we used three values, $\epsilon$= 0.05, 0.22 and 0.3, allocating
nuclei to these categories according to $R_{4/2}$ (see
Fig.~1(right)), and extending these choices to unknown nuclei
using the P-factor \cite{P-factor}.

\begin{figure}
\includegraphics[width=0.47\textwidth]{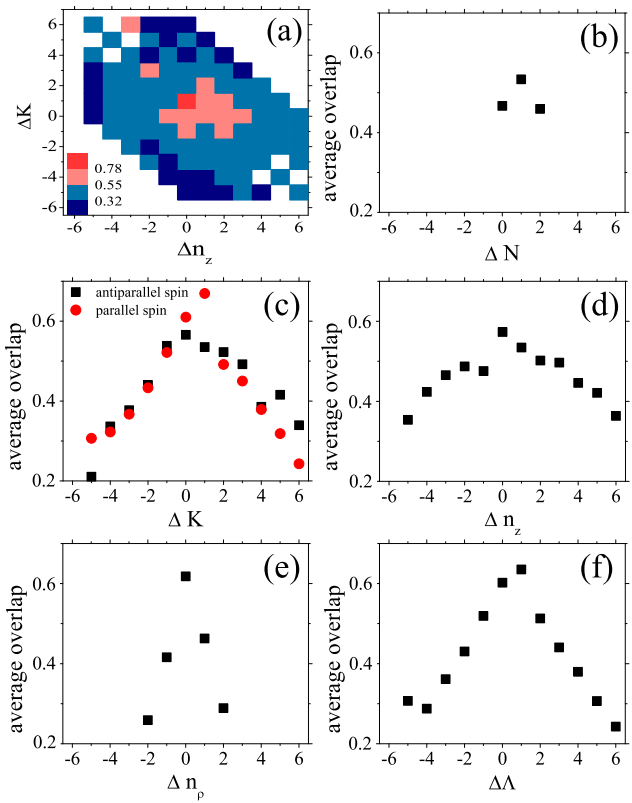}\vspace{0.2cm}
\centering \caption{(Color online) Calculated average spatial
overlaps (for a deformation $\epsilon$= 0.22) for proton and
neutron orbitals in the $Z$= 50-82, $N$= 82-126 region against the
differences (neutron orbit minus proton orbit) in their $K$ and
$n_z$ values in a color code (a). Other panels show average
overlaps as a function of differences [$\Delta$$N$ (b),
$\Delta$$K$ (c), $\Delta$$n_z$ (d), $\Delta$$n_{\rho}$ (e),
$\Delta$$\Lambda$ (f)] in individual Nilsson quantum numbers.}
\end{figure}

\begin{figure}
\includegraphics[width=0.63\textwidth]{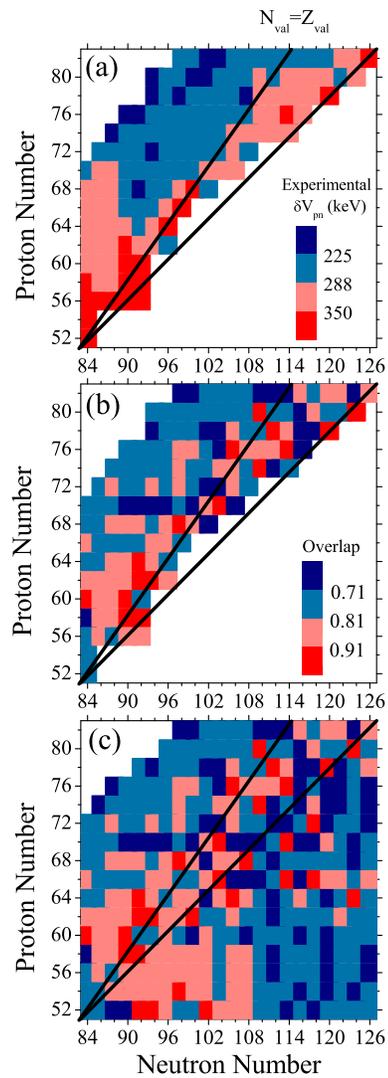}
\centering \caption{(Color online) Color coded empirical
$\delta$$V_{pn}$ values for the $Z$= 50-82 and $N$= 82-126 shells
(a). Large values have redder colors. Similar as (a) but for
calculated overlaps for nuclei where empirical values of
$\delta$$V_{pn}$ are known (b). Calculated overlaps for the full
major shells (excluding nuclei beyond the proton dripline) (c).
The upper (lower) black lines represent $Z_{val}$=$N_{val}$ (equal
fractional filling).}
\end{figure}

It is instructive to look globally at the overlaps. Figure 2 shows
their behavior against correlated differences in $K$ and $n_z$ as
well as against differences in each of the Nilsson quantum
numbers. In panel (a) the overlaps are highest when $\Delta$$K$
and $\Delta$$n_z$ are small, including the 1[000] case involving
proton unique parity orbitals and the case of present interest
0[110]. The overlaps generally fall off for larger $\Delta$$K$ and
$\Delta$$n_z$ values.

\begin{figure*}
\includegraphics[width=0.98\textwidth]{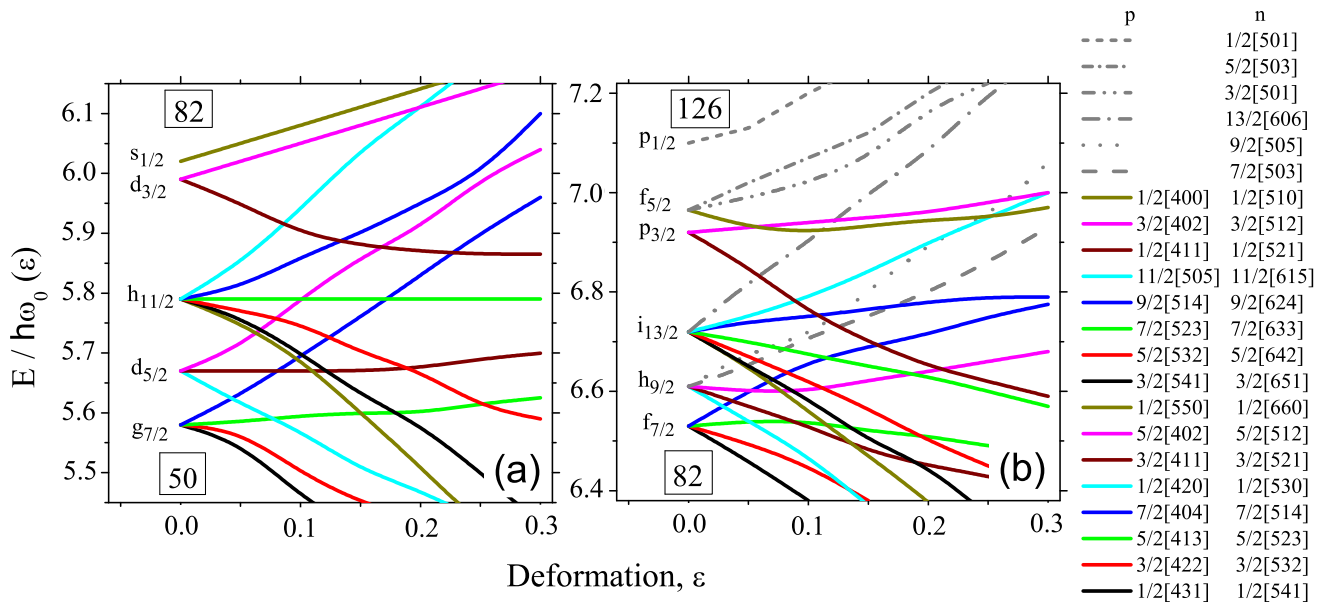}\vspace{-4cm}
\centering \caption{(Color online) Nilsson diagrams for the proton
(a) $Z$= 50-82 and neutron (b) $N$= 82-126 shells. The sequential
filling of $\Delta K$[$\Delta N$,$\Delta{n_z}$,$\Delta
\Lambda$]=0[110] pairs is closely followed for most deformations
in the actual Nilsson diagram as seen by the corresponding color
coding of respective proton and neutron orbitals. Neutron orbitals
without 0[110] proton partners (these have $n_z$= 0) are shown as
black lines in the neutron Nilsson diagram.}
\end{figure*}

However one notes two outlying pink boxes at the upper left in
panel (a). These occur for large values of $\Delta$$K$ (3 and even
6) such as the orbital pair 1/2[431] and 13/2[606] and were at
first rather puzzling. To understand these and the other patterns
we show a further analysis of the overlaps in panels (b)-(f). Each
point is an average over all the overlaps for that value of the
difference in the relevant Nilsson quantum number. In each case,
the overlaps fall off steeply as the particular quantum number
differs by larger and larger amounts in the two orbits, peaking at
a quantum number difference of zero or one (for $\Delta$${N}$ and
$\Delta$${\Lambda}$ - see below). Note that the steepest
dependence is for the $\Delta$n$_{\rho}$ plot at bottom left,
where $\Delta$n$_{\rho}$ is the difference in the number of radial
nodes with n$_{\rho}$= ($N$ $-$ n$_z$ $-$ $\Lambda$)/2. Finally,
the peak at $\Delta$${N}$= +1 is interesting. Given that the
maximum overlaps occur for $\Delta$$n_z$ and $\Delta$n$_{\rho}$=
0, the peak at $\Delta$$N$= +1 implies a corresponding peak at
$\Delta$$\Lambda$= 1 which is indeed seen. We can now understand
the pink boxes with large $\Delta$$K$ in the top left panel (c).
They all correspond to cases of $\Delta$n$_{\rho}$= 0 for which
the large $\Delta$n$_{\rho}$ overlaps compensate for the large
$\Delta$$K$ and $\Delta$$n_z$ values. However, such orbit pairs
form the ground states only in neutron-rich nuclei not currently
accessible.

Figure 3 shows empirical values of $\delta$$V_{pn}$ (a) and our
calculated overlaps (b and c). Overall the agreement is quite good
given the simplicity and parameter free nature of our approach,
and is comparable to that from DFT calculations ~\cite{Mario}. The
results generally show small values far from the diagonal, a
spread out region of large values early in the shells, and large
values near the $Z_{val}$=$N_{val}$ line that shift slightly to
the right of the $Z_{val}$=$N_{val}$ line towards the end of the
shell. A possible reason for this latter behavior will be evident
below. There are occasional pink boxes to the upper left that
disagree with the data. They correspond to very neutron-deficient
isotopes for $Z$$\sim$ 72-76. Note also that the blue box for Pb
at $N$= 124 would be light pink were zero deformation (instead of
0.05) to be used.

Of course, calculated values are not limited to known nuclei.
Figure 3(c) shows overlaps for the full shells. Interestingly,
large overlaps now also appear (as in DFT calculations
\cite{Mario}) in neutron-rich nuclei in the region $Z$$\sim$ 52-64
and $N$$\sim$ 92-108. Here pairs of orbitals, such as 5/2[413]
with 5/2[512] and 1/2[420] with 1/2[521], coupled to $S$= 0, are
filling (near $^{168}$Gd and $^{162}$Nd, respectively), that do
not satisfy 0[110], which implies $S$= 1. Measurement of masses in
these regions, which may be available in the future at FAIR, FRIB,
and RIKEN, would offer important tests of the current ideas.

\section{Implications for the Development of Collectivity and Deformation}

The idea of $p$-$n$ Nilsson orbital pairs related by 0[110] has a
much deeper consequence related to the overall emergence of
collectivity in nuclei. In Fig.~4 we show standard proton (a) and
neutron (b) Nilsson diagrams for this mass region. We first note
that every one of the 16 Nilsson proton orbitals for the entire
shell, including the unique parity orbitals, has a 0[110] neutron
partner. This in itself is perhaps not surprising since the
neutron shell has one additional quantum. However, a closer look
shows a general pattern, not heretofore recognized, namely, that,
these 0[110] pairs fill almost \textit{synchronously} as the
proton and neutron shells fill. This is obvious for small
deformations. For example, one has the successive p-n
combinations: 1/2[431] - 1/2[541]; 3/2[422] - 3/2[532]; 5/2[413] -
5/2[523]; 1/2[420] - 1/2[530]; etc. Since the patterns of up-and
down- sloping orbits and orbit crossings are similar in the two
shells, this synchronous filling of 0[110] combinations
approximately persists even as the deformation increases. For
example, near mid-shell for $\epsilon$ $\sim$0.3, one has,
starting at $Z$= 68 and $N$= 100 (18 valence nucleons each):
7/2[523] - 7/2[633]; 1/2[411] - 1/2[521]; 5/2[402] - 5/2[512];
7/2[404] - 7/2[514]; 9/2[514] - 9/2[624]. Except for one
interchange of adjacent orbits, these continue to fill in highly
overlapping 0[110] combinations even as the deformation changes.
This synchronous filling sequence correlates with, and gives a
microscopic basis to, the empirical phenomenon of enhanced
collectivity along the $Z_{val}$=$N_{val}$ line.

It is only past mid-shell that neutron orbitals occur (6 of 22)
that do not have a 0[110] proton partner. Interestingly, each of
these has $n_z$= 0, that is, oblate orbitals that do not
contribute to prolate deformation. The interspersing of these
rogue $n_z$= 0 orbitals late in the shell interrupts the
$Z_{val}$=$N_{val}$ correlation with maximal $\delta$$V_{pn}$,
leading to shifts in peaks in $\delta$$V_{pn}$ to
$N_{val}$=$Z_{val}$+2 noted earlier (e.g., Hf-W and Lu-Ta region).

\section{A possible new Pseudo-Shell approach to Heavy Nuclei}
The 0[110] correlation is repeatedly encountered from the $sd$
shell to the actinides. This generality may suggest a new coupling
scheme, similar in spirit to the idea of pseudo-SU(3) \cite{Raju,
Draayer}, but different in content. The 50-82 major shell is
formed by the orbits of the $sdg$ oscillator shell, with the
exception that the 1g$_{9/2}$ orbit has escaped into the 28-50
major shell, and is replaced by the 1h$_{11/2}$ orbit, from the
$pfh$ oscillator shell. As a result, the sdg$_{7/2}$h$_{11/2}$
50-82 shell (with the single orbital 11/2[505] left out) can be
considered as an approximate $sdg$ shell by replacing the
1h$_{11/2}$ orbitals by their 0[110] counterpart 1g$_{9/2}$
orbitals. Whereas, in pseudo-SU(3), the entire unique parity orbit
is excised, here only the single, highest $K$, Nilsson orbital is
excluded. The new scheme could simplify symmetry-based shell model
calculations. Instead of two pseudo-SU(3) shells (with SU(3)
subalgebras) plus two shell model single-j shells (not possessing
SU(3) subalgebras), one has just two approximate shells with SU(3)
subalgebras (plus two high-lying high-$K$ single orbitals, which
can often be ignored), thus deriving from the shell model an
approximate SU(3) symmetry for heavy nuclei, at least for
Z$_{val}$$\simeq$N$_{val}$.

As an example, $^{154}$Sm is considered, for which the Nilsson
deformation parameter is $\epsilon\approx 0.95 \beta_2 \approx
0.32$ \cite{Nilsson,Raman}. From Fig.~4 it is clear that 6 of the
12 valence protons occupy normal parity orbitals in the 50-82
shell, while the other 6 occupy 1h$_{11/2}$ orbitals. In addition,
6 of the 10 valence neutrons occupy normal parity orbitals in the
82-126 shell, while the other 4 occupy 1i$_{13/2}$ orbitals.

1) In the pseudo-SU(3) scheme, the 6 protons of normal parity sit
in the (12,0) irreducible representation (irrep) of U(10) (the
pseudo-shell formed within the 50-82 shell \cite{Raju}), while the
other 6 are outside the pseudo-SU(3) symmetry and have to be
treated separately. Similarly, the 6 neutrons of normal parity sit
in the (18,0) irrep of U(15) (the pseudo-shell formed within the
82-126 shell \cite{Raju}), while the other 4 are outside the
pseudo-SU(3) symmetry and are treated separately. Thus, one has a
(30,0) irrep describing the normal parity nucleons, plus 6 protons
in 1h$_{11/2}$ orbitals, plus 4 neutrons in 1i$_{13/2}$ orbitals.

2) In the present coupling scheme, using the same group
theoretical methods as in \cite{Raju}, we see that all 12 valence
protons sit in the (24,0) irrep of U(15) formed by the 50-82 shell
except the high-lying 11/2[505], which plays no role in
$^{154}$Sm, while all 10 neutrons sit in the (30,4) irrep of U(21)
formed by the 82-126 shell except the high-lying 13/2[606], which
also plays no role. Hence, one has a (54,4) irrep for all valence
nucleons in $^{154}$Sm.

To proceed further, one has to choose a Hamiltonian containing, in
addition to the usual quadrupole-quadrupole and angular momentum
terms, SU(3) symmetry preserving third-order and/or fourth-order
terms \cite{Draayer,Berghe}. Work in this direction is in
progress. Finally, the 0[110] proton-neutron pairs considered in
the present work have $S$= 1. The presence of isoscalar $S$= 1
proton-neutron pairs in competition with isovector $S$= 0 nucleon
pairs has long been considered in medium mass nuclei with
$Z$$\simeq$$N$ \cite{Pittel,Simkovic}. The present work suggests
that similar studies for heavy nuclei with $Z_{val}\simeq
N_{val}$.

\section{Conclusions}
New results, for odd-Z nuclei, show a magnified enhancement of the
large empirical values of $p$-$n$ interactions along the
$Z_{val}\simeq N_{val}$ line in a purer form, without the muting
effects of pairing. These enhanced values are closely correlated
with the development of collectivity, shape changes, and the
saturation of deformation.These strong interactions can be simply
understood in terms of parameter-free spatial overlaps of special
pairs of spin-aligned proton and neutron wave functions differing
by single oscillator quanta along the deformation axis. It is
precisely these highly interacting 0[110] pairs that fill almost
synchronously in heavy nuclei, giving a rationale for the way
collectivity develops across major shells. This points to a
possible, complementary, new symmetry-based coupling scheme for
shell model calculations that is more inclusive than existing
schemes.

\section{Acknowledgments}
We are grateful to W.~Nazarewicz, K.~Heyde, A.~Frank and S.~Pittel
for useful discussions. Work supported by the Max-Planck Society,
by the US DOE under Grant No. DE-FG02-91ER-40609 and by the
Scientific Research Projects Coordination Unit of Istanbul
University under Project Numbers 21658 and 26433. R.B.C.
acknowledges support by the Max-Planck-Partner group.

\end{document}